\documentclass[aoas,preprint]{imsart}

\RequirePackage{amsthm,amsmath,amsfonts,amssymb}
\RequirePackage[authoryear]{natbib}
\usepackage{hyperref}
\usepackage{subcaption}
\usepackage{booktabs}
\usepackage{graphicx}
\usepackage{nicefrac}
\usepackage{subcaption}

\startlocaldefs

\endlocaldefs

\begin{document}

\begin{frontmatter}
  \title{Modeling Spatially Obfuscated Street-Crime Data  using  Log-Gaussian Cox Processes on Metric Graphs}
  \runtitle{LGCP for Theft Crimes on Metric Graphs}

  \begin{aug}
    \author[A]{\fnms{Lulu}~\snm{Jiang}\ead[label=e1]{lulu.jiang@kaust.edu.sa}},
    \author[A]{\fnms{David}~\snm{Bolin}\ead[label=e2]{david.bolin@kaust.edu.sa}}
    \address[A]{Computer, Electrical and Mathematical Sciences and Engineering Division, King Abdullah University of Science and Technology \printead[presep ={,\ }]{e1,e2}}
  \end{aug}

    \begin{abstract}
    We develop a log-Gaussian Cox process framework for modelling street-level crime data observed on a road network when the released event locations are spatially obfuscated. Motivated by UK Police street-level crime data, where published coordinates are anonymised proxy locations rather than exact event locations, we address the resulting support mismatch by representing each observation through an aggregated support on the street network. The latent log-intensity is modelled as a Whittle--Mat\'ern Gaussian field defined on a metric graph through an SPDE representation, allowing the crime intensity to vary continuously along streets while respecting the geometry of the road network. We compare the proposed metric-graph aggregated model with two alternatives: a planar point model that treats the released locations as exact points, and a planar aggregated model that accounts for spatial aggregation but ignores the network support. In a simulation study where data are generated on a street network, the metric-graph model provides more accurate parameter recovery and better overall fit than the planar alternative. In the City of London application, the metric-graph model also achieves the best fit across several crime types, including \emph{theft from the person}, \emph{robbery}, \emph{drugs}, and \emph{bicycle theft}. The results further suggest that the relationship between environmental amenities and crime risk varies by crime type, with supermarkets showing the most consistent positive associations. The proposed framework provides a principled approach for analysing network-constrained spatial event data with privacy-protected 
    and imprecise 
    locations.
    \end{abstract}


\end{frontmatter}

\section{Introduction}
The analysis of crime data is an active area of research, motivated by both methodological challenges and high societal relevance.
Existing crime research spans a wide range of methodological approaches. Some studies focus on social structures and individual behaviors, for example using social network analysis  to identify key actors in criminal networks \citep{schwartz2008targeting,schwartz2009using} and agent-based modeling to simulate residential burglary processes \citep{malleson2011using}. Other studies focus on spatio-
temporal dependence and hotspot dynamics, including displacement metrics \citep{bowers2003measuring}, clustering detection methods \citep{jiang2021discovering}, and risk terrain modeling and its variants \citep{andresen2018predicting,connealy2019risk,caplan2020using,drawve2016predictability}. Regression-based approaches and machine learning methods are also widely used to relate crime to socioeconomic and environmental factors \citep{andresen2006spatial,ahmar2018crime,osgood2017poisson,luo2022associating,jung2020spatiotemporal,oh2022evaluation,saxena2019mining}, while Bayesian models have been adopted for dependency modeling and small-area crime analysis \citep{marchant2018applying,quick2018crime}. In addition, spatial and spatio-temporal point process models provide a probabilistic framework for random event data, including Cox processes, marked point processes, and Hawkes-type models \citep{diggle1979parameter,moller2010structured,mohler2014marked,hawkes1971spectra,mohler2011self,reinhart2018self,zhuang2019semiparametric}.

Despite the rich literature on crime analysis, certain crime types, such as theft from the person, robbery, and some forms of street violence, still present a fundamental modeling challenge as their occurrence is strongly constrained by the street network. Event locations are often more naturally represented by roads than by administrative areas or regular grids. Much of the previous analysis, however, ignores the road network and treats such events as points in a homogeneous two-dimensional planar domain, analyzed using Euclidean distance. This is inconsistent with the geometric constraints underlying the event-generating mechanism. \citet{baddeleyAnalysingPointPatterns2021a} explicitly note that, for events confined to roads, it is statistically incorrect to ignore network geometry and analyze the data as an ordinary two-dimensional point pattern, as this may yield spuriously high density values in areas where the road network is denser. \citet{dangeloSelfexcitingPointProcess2024} model crimes as constrained point patterns on linear networks and show that network geometry and network distance definitions are central to model fitting and interpretation. Similarly, \citet{vlad2023analysis} show in a Mexico City crime application that planar kernel density estimation and Euclidean $K$-function tools can produce misleading conclusions for street-network data, for example by assigning probability mass to locations off the network.

Recent developments in stochastic modelling on metric graphs provide an important theoretical and computational foundation for probabilistic modelling on road networks. \citet{moller2024cox} proposed Cox process models on linear networks in which the random intensity is driven by transformed isotropic Gaussian processes, including log-Gaussian Cox processes, interrupted Cox processes and permanental Cox processes. Further, Gaussian Whittle--Mat\'ern fields on compact metric graphs provide a natural extension of Mat\'ern fields from Euclidean domains to network domains \citep{bolin2024gaussian}. Theoretical results on their regularity, numerical approximation, and Markov properties have strengthened the use of these fields as latent Gaussian models on metric graphs \citep{bolin2024regularity,bolin2026markov}. Building on this, log-Gaussian Cox process models on general metric graphs have been developed to enable Bayesian inference for point process data on complex network domains \citep{bolinLogGaussianCoxProcesses2025}. These advances make it possible to model spatial dependence, quantify uncertainty, and estimate covariate effects within a unified Bayesian framework while respecting network structure. 

However, for publicly available crime datasets such as UK police street-level crime data, using one of these metric graph models to get the geometric domain right is not sufficient as these data are released using a snap-point anonymization mechanism, so the published coordinates are not the true crime locations but privacy-protected approximate locations. \citet{tompsonUKOpenSource2015} show that the anonymization procedure used in the UK police street-level crime data can substantially affect spatial accuracy, especially for fine-scale analyses, and can introduce systematic distortions if the data-generation and publication mechanism is ignored. More broadly, the geomasking literature has repeatedly emphasized the trade-off between privacy protection and spatial analytical utility \citep{wangGeomaskingSafeguardGeoprivacy2024}. For statistical crime modeling, this should be handled as a support mismatch problem, where the observed released point is an anonymized proxy, while the underlying event occurred somewhere within a local region of the road network.

We compare three classes of spatial models for this type of spatially obfuscated crime data. The first is a standard Poisson regression, treating the recorded locations as true crime locations and thus ignoring the location obfuscation entirely. The second is an aggregated LGCP, in which the observed counts are assumed to be aggregated over a local spatial support region on $\mathbb{R}^2$.  This approach handles the spatial support mismatch induced by location anonymization. However, the underlying spatial process is still defined on a two-dimensional Euclidean domain and therefore does not capture the geometric constraints imposed by the road network on street crime. The third is a  \emph{metric-graph aggregated LGCP} which we develop in this work. This model incorporates two key features simultaneously. First, the occurrence of street crime is inherently constrained by the structure of the road network by assuming that the true crime locations follow an LGCP on the street network, defined via a Whittle--Mat\'ern field on the metric graph representing the street network. Second, the observed counts are assumed to be aggregated over local support regions on the road network. 

Our results show that explicitly modelling both the network-constrained structure of crime data and the aggregation induced by location anonymisation provides clear advantages. In the simulation study, where the data-generating process is defined on a street network, the proposed metric-graph aggregated LGCP provides more accurate estimation of the key model parameters than the planar alternatives. In particular, by defining both the latent process and the aggregation supports on the metric graph, the proposed model reduces bias in the fixed-effect estimates and gives more reliable inference for the spatial dependence parameters, including the range and marginal variation of the latent field. In the City of London application, the proposed model achieves the best overall fit for all four crime types and yields interpretable covariate effects based on network distances to local amenities. These results indicate that the proposed framework is particularly well suited to privacy-protected street-crime data, since it is consistent with the location anonymisation mechanism.

This paper is organized as follows. Section~\ref{sec2} describes the crime data and the anonymization mechanism. Section~\ref{sec3} introduces the methodological background, including metric graphs, Gaussian Whittle--Mat\'ern fields, and log-Gaussian Cox process, and the construction of integration points on metric graphs. Section~\ref{sec4} introduces the proposed area-aggregated LGCP while Section~\ref{sec5} introduces the two alternative models. A simulation study is presented in Section~\ref{sec6} and  Section~\ref{sec7} applies the proposed methodology to crime data from the City of London. Section~\ref{sec8} concludes with a discussion.

\section{Data}
\label{sec2}
We download crime and policing data from \url{https://data.police.uk/}, which provides open downloads and API access for England, Wales, and Northern Ireland, including monthly updated street-level crime records and anti-social behaviour (ASB) incidents. For privacy protection, the platform systematically anonymises several sensitive fields: temporal information is retained only at the year--month level, while spatial locations are obfuscated through an anonymous ``snap point'' mechanism. Locations in the street-level crime data are only approximations of where crimes actually occurred, rather than exact coordinates \citep{datacrimeuk}. The snap point mechanism maintains a list of anonymous map points and uses these as proxy locations in the publicly released data. This list was first created in 2012 and refreshed in 2022 using updated geographic data sources. Its construction combines several types of spatial information. First, the centre point of each road in England and Wales was extracted from the Ordnance Survey Locator dataset to form an initial set of candidate anonymous map points. Second, these candidates were augmented with locally relevant points of interest from PointX, a national point-of-interest database containing geocoded records of educational institutions, leisure services, retail and commercial premises, and other landscape features \citep{pointx}. These additional points, such as parks, airports, shopping centres, and other public places, were used to improve coverage in non-residential areas. Third, the number of postal addresses within the catchment area (defined through a Voronoi tessellation) of each candidate point was calculated using the Ordnance Survey Address-Point dataset, and candidate points with between one and seven postal addresses were removed to satisfy the privacy requirement. Finally, the remaining points were reviewed by police forces, and a small number of additions and deletions were made based on local feedback to improve their local representativeness. After this procedure, each snap point is located on a street centreline, in a public place, or at a commercial landmark, and its catchment area contains either at least eight postal addresses or no postal addresses at all. When data are uploaded by police forces, the true coordinates of each crime are replaced by the coordinates of the nearest anonymous point. If the nearest snap point is more than 20 km away, coordinates of zero are assigned, so that the record is not shown on the subsequent public crime map \citep{tompsonUKOpenSource2015}. 

\begin{figure}[t]
  \centering
  \includegraphics[width=0.8\linewidth]{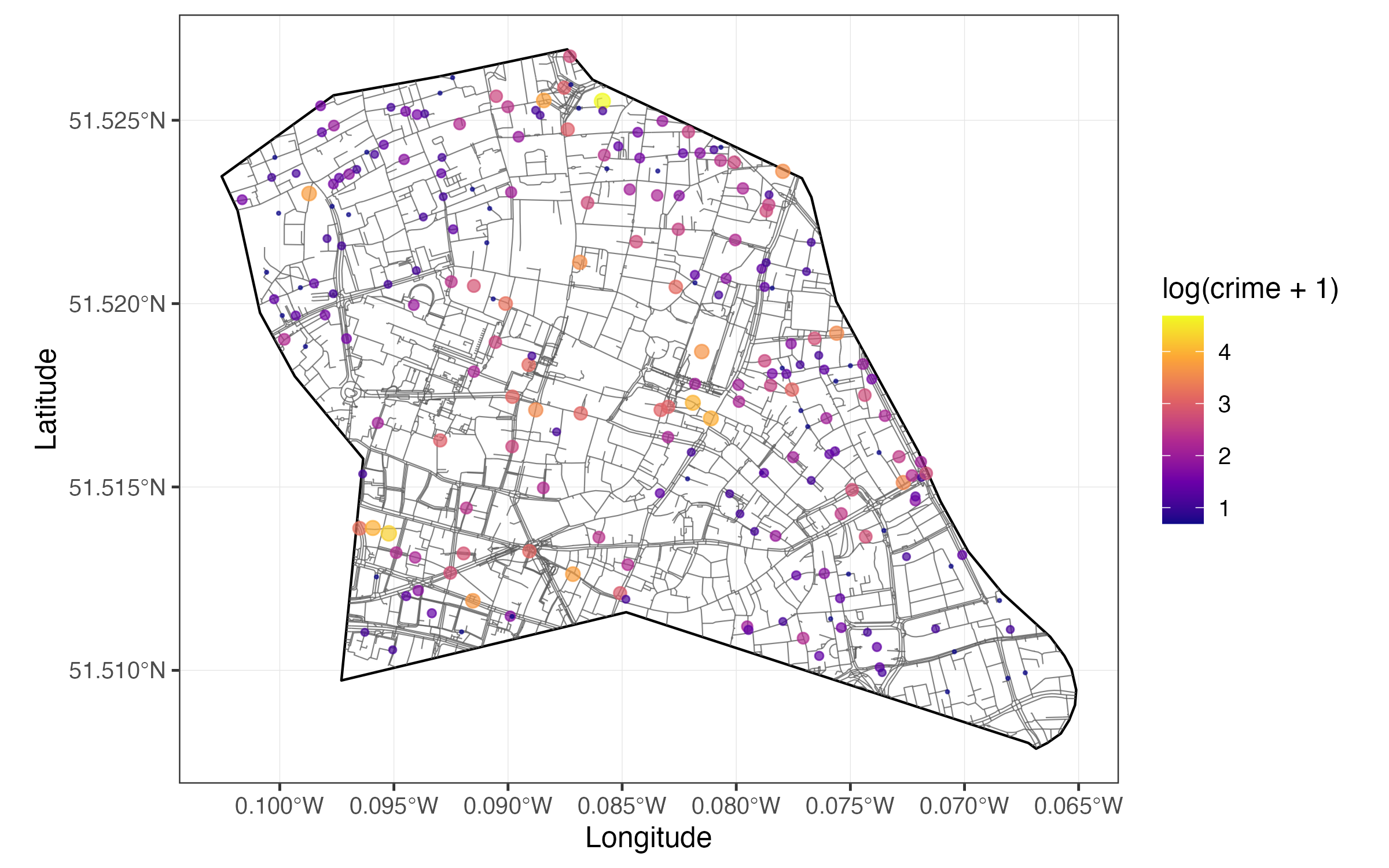}
  \caption{Aggregated spatial distribution of theft from the person in the City of London in 2023 based on anonymised street-level UK Police data. Each point represents an assigned snap-point location, and the displayed values correspond to the logarithmically transformed crime counts, $\log(\mathrm{crime}+1)$, to improve visual interpretability.}
  \label{fig:theft_data_2023}
  % \db{ADD THE METRIC GRAPH TO THE PLOT. THE TITLE OF THE COLORBAR IS STRANGE}
\end{figure}

This type of spatial anonymisation is not unique to \texttt{data.police.uk}; rather, it is a standard privacy-preserving practice in public crime data systems. Although the exact implementation varies across jurisdictions, many crime databases reduce spatial precision before publication. For example, the Chicago open crime data report addresses only at the block level and do not identify specific locations \citep{chicago_crimes_dataset}; the Los Angeles crime data provide address fields only to the nearest hundred block \citep{datagov_lapd_crime_data_2020_2024}; and the San Francisco Police Department incident-report data map incident locations to nearby intersections to ensure anonymity \citep{datasf_sfpd_incident_reports_2018_present}. Therefore, the snap point mechanism used by \texttt{data.police.uk} can be understood as one specific implementation of a broader practice in public crime data release, namely protecting privacy by reducing spatial precision, rather than as an exceptional data-processing procedure. As mentioned in the introduction, treating the released coordinates as exact event locations under such anonymisation can therefore introduce bias, motivating modelling approaches that are consistent with the data-generation and release mechanism. This is the main motivation for developing the models in this work. As an illustrative example, we take the City of London as our study region. We construct the street network from OpenStreetMap road data and represent it as a metric graph. The resulting network contains 6{,}909 intersections (vertices) and 8{,}230 road segments (edges). Segment lengths range from 10\,m to 298\,m, and the total network length is 163.9\,km. To characterise environmental exposure, we use as covariates the shortest-path distance along the street network from each location to the nearest amenity of a given type. We download street-level records from the UK Police API for the period January 2023 to December 2025, retaining only observations within the study boundary. To exploit temporal repetition while controlling computational cost, we aggregate months into annual groups and treat each year as a replicate. Within each year, we count events sharing the same assigned location, yielding replicate-specific counts aggregated by location. Figure~\ref{fig:theft_data_2023} shows the aggregated spatial distribution of theft from the person in 2023. 

\section{Background}
\label{sec3}
\subsection{Metric graphs}
A metric graph provides a natural representation of spatial domains whose geometry is constrained by a network, such as street, traffic, or river systems. Let 
\(\Gamma=(\mathcal{V},\mathcal{E})\) be a compact undirected metric graph, where \(\mathcal{V}\) is a finite set of vertices and \(\mathcal{E}\) is a finite set of edges. Each edge \(e\in\mathcal{E}\) is associated with a finite positive length \(l_e\in(0,\infty)\) and is parameterized by arclength \(t\in[0,l_e]\). Hence, any location on the graph can be written as \(s=(e,t)\in\Gamma\), where \(e\) identifies the edge and \(t\) denotes the distance from one endpoint of that edge. Since \(\Gamma\) is assumed to be connected, the natural distance between two locations \(s,s'\in\Gamma\) is the geodesic distance \(d(s,s')\), defined as the length of the shortest continuous path along the edges connecting them. On a metric graph, functions, covariates, point patterns, and random fields can be defined continuously along both vertices and edge interiors \citep{bolin2024gaussian,bolin2026statistical,bolin2025new}. Figure \ref{fig:theft_data_2023} shows an example of such a metric graph constructed from the street network in the study region. The grey line segments represent the edges of the graph, while the black boundary indicates the spatial domain considered in the analysis.  In the subsequent modelling framework, the latent spatial process is therefore defined on \(\Gamma\), allowing prediction and inference at arbitrary locations along the street network rather than only at the vertices.

\subsection{Gaussian Whittle--Mat\'ern fields}
Let $\mathcal{D}\subset \mathbb{R}^d$ be a bounded Euclidean domain. A Gaussian Whittle--Mat\'ern field on $\mathcal{D}$ is defined as a solution to the stochastic partial differential equation (SPDE)
\begin{equation}
(\kappa^2-\Delta)^{\alpha/2}(\tau u)=\mathcal{W},
\qquad s\in \mathcal{D},
\label{eq:spde_euclidean}
\end{equation}
where $\Delta$ is the Laplacian equipped with Neumann boundary conditions, $\mathcal{W}$ is Gaussian white noise, $\tau>0$ is a scaling parameter, $\kappa>0$ a range parameter and $\alpha=\nu+d/2$ for a smoothness parameter $\nu>0$. The motivation for these Gaussian processes is that if $\mathcal{D} = \mathbb{R}^d$, the corresponding Gaussian process has the Mat\'ern covariance function
\begin{equation*}
\mathrm{Cov}\{u(s),u(s')\}
=
\sigma^2
\frac{2^{1-\nu}}{\Gamma(\nu)}
(\kappa \|s-s'\|)^\nu
K_\nu(\kappa \|s-s'\|),
\label{eq:matern_cov}
\end{equation*}
where $\sigma^2$ is the marginal variance, determined by $\tau,\kappa,\alpha$, and
$K_\nu(\cdot)$ is the modified Bessel function of the second kind \citep{whittle1954stationary,whittle1963stochastic}.
This SPDE representation of Gaussian Mat\'ern fields is important because it links them to Gaussian Markov random fields, yielding sparse precision matrices and therefore computationally efficient inference \citep{lindgren2011explicit}. The idea is to approximate the solution of the SPDE by a finite element expansion 
\begin{equation*}
u_h(s)=\sum_{j=1}^{n_h} w_j \psi_j(s),
\label{eq:fem_expansion_euclidean}
\end{equation*}
with piecewise linear hat basis functions $\{\psi_j\}$ defined on a triangulation of the domain. Let $C$ and $G$ denote the sparse mass and stiffness matrices, with elements 
$C_{ij}=\langle \psi_i,\psi_j\rangle$ and 
$G_{ij}=\langle \nabla \psi_i,\nabla \psi_j\rangle$,
and define $K=\kappa^2 C + G$. When $\alpha=1$, the finite element approximation leads directly to a Gaussian Markov random field with sparse precision matrix
$Q = K$.
When ${\alpha=2}$, the approximate field has precision matrix
$Q = K^\top C^{-1} K$.
Since $C^{-1}$ is generally dense, a lumped diagonal approximation of $C$ is often used in practice to recover sparsity. This approximation enables efficient Bayesian inference within latent Gaussian models through the INLA framework of \cite{rue2009approximate}.

\cite{bolin2024gaussian} extended the SPDE construction of
Whittle--Mat\'ern fields to compact metric graphs. 
In this case, the Laplacian is replaced by the Kirchhoff Laplacian $\Delta_\Gamma$, which acts as the second derivative in the interior of each edge, and which is equipped with  the Kirchhoff vertex conditions
\begin{equation}
f \text{ is continuous on } \Gamma,
\qquad
\sum_{e\in\mathcal{E}_v}\partial_e f(v)=0,
\quad v\in\mathcal{V},
\label{eq:kirchhoff_condition}
\end{equation}
where $\mathcal{E}_v$ denotes the set of edges incident to $v$ and
$\partial_e f(v)$ is the directional derivative along edge $e$ away
from $v$.  A Gaussian Whittle--Mat\'ern field on $\Gamma$ is then defined as the
solution of
\begin{equation}
(\kappa^2-\Delta_\Gamma)^{\alpha/2}(\tau u)
=
\mathcal{W},
\qquad \text{on }\Gamma,
\label{eq:spde_metric_graph}
\end{equation}
where $\mathcal{W}$ is Gaussian white noise, $\kappa>0$ controls the
spatial dependence scale, $\tau>0$ is a scaling parameter, and
$\alpha=\nu+1/2$ since the edges of the metric graph are
one-dimensional. In this work, we restrict attention to $\alpha=1$, corresponding to
$\nu=1/2$. In this case, the field evaluated at the vertices has
an exact Gaussian Markov random field representation
\citep{bolin2026statistical}. Let $\mathbf{u}_{\mathcal V}
=
\bigl(
u(v_1),\ldots,u(v_{|\mathcal V|})
\bigr)^\top$
Then
$\mathbf{u}_{\mathcal V}
\sim
\mathcal N(\mathbf 0,Q_\Gamma^{-1})$
where, by Corollary~2 of \citet{bolin2026statistical}, the entries of
the precision matrix $Q_\Gamma$ are
\begin{equation}
(Q_\Gamma)_{ij}
=
2\kappa\tau^2
\begin{cases}
\displaystyle
\sum_{e\in\mathcal E_{v_i}}
\left[
\frac{1}{2}
\frac{1+\exp(-2\kappa\ell_e)}
     {1-\exp(-2\kappa\ell_e)}
\mathbb I(\underline e\neq\overline e)
+
\tanh\left(\frac{\kappa\ell_e}{2}\right)
\mathbb I(\underline e=\overline e)
\right],
& i=j,
\\[5mm]
\displaystyle
-\sum_{e\in\mathcal E_{v_i}\cap\mathcal E_{v_j}}
\frac{\exp(-\kappa\ell_e)}
     {1-\exp(-2\kappa\ell_e)},
& i\neq j,
\end{cases}
\label{eq:graph_precision}
\end{equation}
where $\underline e$ and $\overline e$ denote the two endpoints of the 
edge $e$. In particular, $(Q_\Gamma)_{ij}=0$ whenever $v_i$ and
$v_j$ are not connected by an edge, so $Q_\Gamma$ inherits its sparse
structure directly from the graph topology. The same construction can be used to evaluate the field at arbitrary locations in the interiors of the edges. Since adding or removing
degree-two vertices does not change the Whittle--Mat\'ern field
\citep{bolin2024gaussian}, any locations of interest can be inserted as
degree-two vertices by subdividing the corresponding edges. The
precision matrix for the augmented graph is then obtained directly
from \eqref{eq:graph_precision}. Thus, for $\alpha=1$, the
finite-dimensional distribution of the continuous field at any
prescribed set of locations can be evaluated exactly through the
sparse GMRF representation, without requiring a finite element
approximation. In this paper, the latent field is implemented using
\texttt{graph\_spde} in the \texttt{MetricGraph} framework with
$\alpha=1$. We parameterize the model through $\sigma
= \nicefrac{1}{(\tau\sqrt{2\kappa})}$ and 
$\rho = \nicefrac{2}{\kappa}$, 
where $\sigma$ is a scale parameter corresponding to the
one-dimensional Mat\'ern standard deviation and $\rho$ is the associated practical correlation range.

\subsection{Log-Gaussian Cox processes}
A log-Gaussian Cox process (LGCP) on a spatial domain \(\Omega\), such as a bounded Euclidean domain or a metric graph, is a Cox process whose random intensity is the exponential of a Gaussian process on $\Omega$. 
If \(Y=\{s_1,\ldots,s_N\}\subset\Omega\) is an observed point pattern and
\(\lambda(s)=\exp\{Z(s)\}\), where \(Z(s)\) is a Gaussian process, the conditional likelihood is
\begin{equation*}
\pi(Y\mid \lambda)
=
\exp\left\{
|\Omega|-\int_{\Omega}\lambda(s)\,ds
\right\}
\prod_{s_i\in Y}\lambda(s_i),
\end{equation*}
The main computational challenge is the integral of the random intensity. Classical approaches often approximate the point process by counts on a fine grid, but this may be inefficient because the grid is used both to approximate the latent Gaussian field and to approximate the event locations. The ``going off grid'' approach of \citet{simpson2016going} instead keeps the observed events at their original continuous locations and approximates only the Gaussian process through a finite element approximation and the integral in the likelihood by numerical quadrature. This gives a likelihood approximation that can be written in Poisson form and fitted within latent Gaussian inference frameworks such as INLA. In this paper, we consider LGCPs on a compact metric graph \(\Gamma\), with random intensity $\Lambda(s)=\exp\{\eta(s)\}$. We specify the latent log-intensity as
\begin{equation}
\eta(s)=\beta_0+\mathbf{z}(s)^\top\boldsymbol{\beta}+u(s),
\label{eq:linear_predictor_graph}
\end{equation}
where \(\beta_0\) is an intercept, \(\mathbf{z}(s)\) is a vector of graph-based covariates, \(\boldsymbol{\beta}\) is the corresponding coefficient vector, and \(u(s)\) is a Whittle--Mat\'ern Gaussian process with $\alpha=1$. For an observed point pattern \(Y=\{s_1,\ldots,s_N\}\subset\Gamma\), the conditional likelihood is
\begin{equation}
\pi(Y\mid \eta)
=
\exp\left\{
|\Gamma|
-\int_{\Gamma}\exp\{\eta(s)\}\,ds
\right\}
\prod_{i=1}^N \exp\{\eta(s_i)\}.
\label{eq:lgcp_likelihood_graph}
\end{equation}
Following \citet{bolinLogGaussianCoxProcesses2025}, the graph integral in \eqref{eq:lgcp_likelihood_graph} is approximated by quadrature points \(\{\tilde{s}_j\}_{j=1}^p\) and weights \(\{\tilde{a}_j\}_{j=1}^p\) placed along the graph,
\begin{equation*}
\int_{\Gamma}\exp\{\eta(s)\}\,ds
\approx
\sum_{j=1}^p \tilde{a}_j\exp\{\eta(\tilde{s}_j)\}.
\label{eq:lgcp_quadrature_graph}
\end{equation*}
This yields the approximate log-likelihood
\begin{equation*}
\log \pi(Y\mid \eta)
\approx
|\Gamma|
-\sum_{j=1}^p \tilde{a}_j\exp\{\eta(\tilde{s}_j)\}
+
\sum_{i=1}^N \eta(s_i),
\label{eq:offgrid_likelihood_graph}
\end{equation*}
which can be evaluated without a finite element approximation of the Whittle--Mat\'ern field if $\alpha$ is an integer. This leads to an inference scheme for LGCPs on general compact metric graphs which is implementable in the R-INLA software.

\section{Area-aggregated Log-Gaussian Cox process}
\label{sec4}
In this section, we formulate an area-aggregated LGCP on a compact metric graph \(\Gamma\), to handle the spatial obfuscation issue with crime data. 

\subsection{Definition and inference}
Suppose that the spatial domain is partitioned into $M$ disjoint aggregation areas $\{B_i\}_{i=1}^M$, corresponding to $M$ snap points for the crime data. Let $A_i = B_i\cap\Gamma$ be the area intersected with the metric graph, which thus consists of one or more line segments in the graph, see Figure \ref{fig:spatial_vs_mg_merged_ips} for an illustration. 
% \db{FIGURE X}
We assume that the true crime locations follow an LGCP $X$ on $\Gamma$ with a log-intensity of the form \eqref{eq:linear_predictor_graph}, where $u$ is a Gaussian Whittle--Mat\'ern field on $\Gamma$ in  \eqref{eq:spde_metric_graph} with $\alpha=1$. Let $N(A_i)=|X\cap A_i|$ denote the observed number of events in area $A_i$. Conditional on the latent field $\eta$, the Cox process reduces to an inhomogeneous Poisson process, so that
\begin{equation}
N(A_i)\mid \eta \stackrel{\text{ind}}{\sim} \mathrm{Poisson}(\mu_i),
\qquad
\mu_i=\int_{A_i}\exp\{\eta(s)\}\,ds,
\quad i=1,\ldots,M.
\label{eq:poisson_area}
\end{equation}
Thus, the joint likelihood for the aggregated counts $\mathbf{N}=(N(A_1),\ldots,N(A_M))$ is
\begin{equation*}
\pi(\mathbf{N}\mid \eta)
\propto
\prod_{i=1}^M
\left(
\int_{A_i} e^{\eta(s)}\,ds
\right)^{N(A_i)}
\exp\left(
-\sum_{i=1}^M \int_{A_i} e^{\eta(s)}\,ds
\right).
\label{eq:lgcp_likelihood}
\end{equation*}
This likelihood is not tractable due to the integrals, and we therefore follow \citet{bolinLogGaussianCoxProcesses2025} and approximate the integrals using numerical integration as explained in the next subsection.

\subsection{Numerical integration on metric graph}
\label{Int}
To evaluate the aggregated intensity as defined in \eqref{eq:poisson_area}, we need a numerical integration scheme on the metric graph. Since each aggregation area $A_i \subset \Gamma$ may consist of several line segments across multiple edges, the integral is interpreted as a line integral over the corresponding subgraph. Hence,
\begin{equation*}
\int_{A_i} f(s)\,ds
=
\sum_{e\in\mathcal{E}} \int_{A_i\cap e} f((e,t))\,dt,
\label{eq:graph_line_integral}
\end{equation*}
for any integrable function $f$ defined on $\Gamma$. We approximate these edgewise integrals using Simpson's rule. For each edge $e$, let
$\{t_k\}_{k=0}^{K}$
be a collection of equally spaced points along the edge. Step size $h_e$ is the maximum distance between adjacent integration points. For a given aggregation area $A_i$, define
$\{t_k^{A_i\cap e}\}_{k=0}^{\tilde K}$
as the grid points lying in the intersection $A_i\cap e$ and $\tilde K$ is an even number. Then the integral over $A_i\cap e$ is approximated by
\begin{equation*}
\int_{A_i\cap e} f((e,t))\,dt
\approx
\frac{h_e}{3}
\left[
f(t_0^{A_i\cap e})
+
4\sum_{k=1}^{\tilde K/2} f(t_{2k-1}^{A_i\cap e})
+
2\sum_{k=1}^{\tilde K/2-1} f(t_{2k}^{A_i\cap e})
+
f(t_{\tilde K}^{A_i\cap e})
\right].
\end{equation*}
Summing over all edges intersecting $A_i$ yields
\begin{equation*}
\int_{A_i} f(s)\,ds
\approx
\sum_{e\in\mathcal{E}}
\frac{h_e}{3}
\left[
f(t_0^{A_i\cap e})
+
4\sum_{k=1}^{\tilde K/2} f(t_{2k-1}^{A_i\cap e})
+
2\sum_{k=1}^{\tilde K/2-1} f(t_{2k}^{A_i\cap e})
+
f(t_{\tilde K}^{A_i\cap e})
\right].
\end{equation*}
Taking $f(s)=\exp\{\eta(s)\}$, the Poisson mean for each block is approximated by
\begin{align*}
\mu_i
=
\int_{A_i}\exp\{\eta(s)\}\,ds 
&\approx
\sum_{e\in\mathcal{E}}
\frac{h_e}{3}
\Big[
\exp\{\eta((e, t_0^{A_i\cap e}))\}
+
4\sum_{k=1}^{\tilde K/2} \exp\{\eta((e, t_{2k-1}^{A_i\cap e}))\} \\
&\qquad
+
2\sum_{k=1}^{\tilde K/2-1} \exp\{\eta((e, t_{2k}^{A_i\cap e}))\}
+
\exp\{\eta((e, t_{\tilde K}^{A_i\cap e}))\}
\Big].
\end{align*}

\subsection{Full likelihood approximation}
We now describe the likelihood approximation used for the main
metric-graph model with replicates. Let $r=1,\ldots,R$ index the
temporal replicates, and let $\{A_i^{MG}\}_{i=1}^{M}$ denote the
graph-based aggregation blocks on the metric graph $\Gamma$. The
observed response $Y_{ir}$ is the number of events in block
$A_i^{MG}$ during replicate $r$. Conditional on the latent intensity,
\begin{equation*}
Y_{ir}\mid \mu_{ir}^{MG}
\sim
\mathrm{Poisson}\!\left(\mu_{ir}^{MG}\right),
\qquad
i=1,\ldots,M,\quad r=1,\ldots,R,
\label{eq:mg_obs_model}
\end{equation*}
where
\begin{equation}
\mu_{ir}^{MG}
=
\int_{A_i^{MG}}
\exp\{\eta_r(s)\}\,ds,
\qquad s\in\Gamma.
\label{eq:mg_mean_integral}
\end{equation}
The latent log-intensity is specified as
\begin{equation}
\eta_r(s)
=
\beta_0
+
\mathbf{z}(s)^\top\boldsymbol{\beta}
+
u_r(s),
\label{eq:mg_eta_replicate}
\end{equation}
where $\mathbf{z}(s)$ is a vector of graph-based covariates and
$u_r(s)$ is a replicate-specific Gaussian Whittle--Mat\'ern field on
$\Gamma$. Conditional on the common hyperparameters, the fields
$u_1,\ldots,u_R$ are independent replicates of the same
Whittle--Mat\'ern model. For $\alpha=1$, the latent field is represented exactly through the
sparse GMRF construction described above. In particular, the numerical
integration points can be inserted into the graph as degree-two
vertices without changing the underlying Whittle--Mat\'ern field
\citep{bolin2024gaussian}. Let $\Gamma^{*}
=
(\mathcal V^{*},\mathcal E^{*})$
denote the augmented graph containing the original graph vertices and
all integration points required for the likelihood approximation, and
write $\mathbf{x}_r=\bigl(u_r(v_1^{*}),\ldots,u_r(v_{n_*}^{*})\bigr)^\top$, where $n_* = |\mathcal V^{*}|$. Then $\mathbf{x}_r\mid\boldsymbol{\theta}\sim\mathcal N\left(\mathbf 0,Q_{\Gamma^{*}}(\boldsymbol{\theta})^{-1}\right)$ for $r=1,\ldots,R$, where $Q_{\Gamma^{*}}(\boldsymbol{\theta})$ is the exact sparse
precision matrix obtained from \eqref{eq:graph_precision} for the
augmented graph. The integral in \eqref{eq:mg_mean_integral} is approximated using the
numerical quadrature scheme described in Section~\ref{Int}. Let
$s_{ij}$ denote the $j$th integration point in $A_i^{MG}$ and
$w_{ij}$ its associated quadrature weight. Then
\begin{equation}
\mu_{ir}^{MG}
\approx
\sum_{j=1}^{m_i}
w_{ij}
\exp\{\eta_r(s_{ij})\}.
\label{eq:mg_mean_quad}
\end{equation}
Since each $s_{ij}$ is included as a vertex of $\Gamma^{*}$, let
$q(i,j)$ denote its corresponding index in $\mathcal V^{*}$. The
linear predictor at the integration points is therefore
\begin{equation*}
\eta_r(s_{ij})
=
\beta_0
+
\mathbf{z}(s_{ij})^\top\boldsymbol{\beta}
+
x_{q(i,j),r}.
\label{eq:eta_quad_gmrf}
\end{equation*}
The latent field value at each integration point is directly
a component of the exact GMRF on the augmented graph. The likelihood contribution from all graph blocks and replicates is
\begin{equation*}
\pi
\left(
\mathbf Y
\mid
\boldsymbol{\beta},
\{\mathbf{x}_r\}_{r=1}^{R},
\boldsymbol{\theta}
\right)
=
\prod_{r=1}^{R}
\prod_{i=1}^{M}
\frac{
\left(\mu_{ir}^{MG}\right)^{Y_{ir}}
\exp\{-\mu_{ir}^{MG}\}
}{
Y_{ir}!
},
\label{eq:full_poisson_likelihood}
\end{equation*}
where $\mu_{ir}^{MG}$ is evaluated using the quadrature approximation
in \eqref{eq:mg_mean_quad}. Together with the Gaussian distributions
of the replicate-specific latent fields, the joint density of the
observations and latent fields is
\begin{equation*}
\begin{split}
&\pi
\left(
\mathbf Y,
\{\mathbf{x}_r\}_{r=1}^{R}
\mid
\boldsymbol{\beta},
\boldsymbol{\theta}
\right)
\\
&\qquad =
\prod_{r=1}^{R}
\left[
\pi(\mathbf{x}_r\mid\boldsymbol{\theta})
\prod_{i=1}^{M}
\frac{
\left(\mu_{ir}^{MG}\right)^{Y_{ir}}
\exp\{-\mu_{ir}^{MG}\}
}{
Y_{ir}!
}
\right].
\end{split}
\label{eq:joint_likelihood_latent}
\end{equation*}
In practice, the integration points and corresponding quadrature
weights are constructed on the metric graph as described in
Section~\ref{Int}, and \texttt{inlabru} is used to construct the
Poisson likelihood based on \eqref{eq:mg_mean_quad} and to perform
Bayesian inference using INLA
\citep{MetricGraph2023,lindgren2024inlabru,rue2009approximate}.

\subsection{Prior specification}\label{sec:prior}
The fixed effects are assigned Gaussian priors with variance \(10\),
$\beta_k \sim N(0,10)$, for 
 $k=0,\dots,p$.
We parametrize the Gaussian process through 
$\theta_1=\log(\sigma)$, 
and
$\theta_2=\log(\rho)$.
% , where \(\sigma\) is the marginal standard deviation and \(\rho\) is the range parameter on the metric graph. 
Here, $\sigma = 1/(\tau\sqrt{2\kappa})$ determines the marginal scale of the latent Gaussian field $u(s)$, while the range parameter $\rho = 2/\kappa$ governs its spatial dependence and is defined with respect to the geodesic distance on $\Gamma$. We assign independent Gaussian priors on the log-scale hyperparameters, 
$\theta_1 \sim \mathcal{N}(\mu_\sigma, s_\sigma^2)$,
and $\theta_2 \sim \mathcal{N}(\mu_\rho, s_\rho^2)$. The prior median of \(\sigma\) is set to \(1\), with $\mu_\sigma=\log(1)$, $\mu_\rho = \log(0.05\,\mathrm{km})$ and 
\begin{equation*}
s_\rho=
\frac{\mu_\rho-\log(0.01)}{\Phi^{-1}(1-0.01)}, 
\qquad
s_\sigma=
\frac{\log(5)-\mu_\sigma}{\Phi^{-1}(1-0.01)}.
\label{eq:mg_sigma_prior_param}
\end{equation*}
With these choices, we have a prior \(P(\rho < 0.01\,\mathrm{km}) = 0.01\) and \(P(\sigma > 5) = 0.01\).

%\subsection{Observation model}
%The response considered in this paper is a count variable obtained by aggregating information from an underlying point pattern. For such data, a Poisson model provides a natural baseline for modelling non-negative integer-valued outcomes. Therefore, all three models are fitted using the \texttt{poisson} likelihood in \texttt{inlabru}. Specifically, conditional on a mean parameter $\mu$, the response $Y$ follows a Poisson distribution with
%\begin{equation*}
%\Pr(Y=y \mid \mu)
%=
%\frac{\mu^y e^{-\mu}}{y!},
%\qquad y=0,1,2,\dots .
%\end{equation*}
%The three models differ not in the observation distribution itself, but in how the mean parameter $\mu$ is constructed.

\section{Alternative models}
\label{sec5}
To assess the effect of modelling the crime process directly on the street network, we compare the metric-graph aggregated model described in Section~\ref{sec4} with two simpler planar alternatives: a point-referenced model (Point) and a planar aggregated model (Spatial). These alternatives differ in the spatial support on which the response is represented. The Point model treats the observed count as attached to a point location, while the Spatial model treats it as an aggregated count over a planar areal unit. This comparison follows the general idea that observations with different spatial supports should be represented through likelihood contributions defined on their corresponding supports, as in recent metric-graph work on joint modelling of point-referenced and line-referenced data \citep{lilleborge2025joint}. Let \(r=1,\ldots,R\) index the temporal replicates. Both alternative models are formulated as latent Gaussian models with Poisson likelihoods. The Point and Spatial models use the same priors as the MG model, specified in Section~\ref{sec:prior}. Specifically, the intercept and covariate coefficients are assigned Gaussian priors and the Mat\'ern SPDE latent fields are also parameterised by
$\theta=(\log\sigma,\log\rho)$,
with Gaussian priors calibrated in the same way as for the MG model.

\subsection{Point model}
The simplest alternative is to treat the count attached to each observed location as a point-referenced response. Let \(s_i\in D\subset\mathbb{R}^2\) denote the \(i\)th observed crime location, and let \(Y_{ir}\) be the count attached to location \(s_i\) in replicate \(r\). The Point model is defined as
\begin{equation*}
Y_{ir}\mid \mu_{ir}^{\mathrm{Point}}
\sim
\mathrm{Poisson}\!\left(\mu_{ir}^{\mathrm{Point}}\right), \qquad 
\mu_{ir}^{\mathrm{Point}}
=
\exp\{\eta_r^{\mathrm{Point}}(s_i)\}.
\label{eq:mu_point_new}
\end{equation*}
The latent predictor is evaluated directly at the observed location:
\begin{equation*}
\eta_r^{\mathrm{Point}}(s_i)
=
\beta_0
+
\sum_{k=1}^{p}\beta_k z_k^{\mathrm{Point}}(s_i)
+
u_r^{\mathrm{Point}}(s_i).
\label{eq:latent_predictor_point_new}
\end{equation*}
Here, \(z_k^{\mathrm{Point}}(s_i) = \exp\left\{-d_k^{\mathrm{Point}}(s_i)\right\}\) 
is the covariate associated with amenity type \(k\), where  \(d_k^{\mathrm{Point}}(s_i)\) is the Euclidean distance from \(s_i\) to the nearest amenity of type \(k\). 
The latent field \(u_r^{\mathrm{Point}}(\cdot)\) is a zero-mean Gaussian Whittle--Mat\'ern field on the planar domain \(D\), represented through the SPDE approach on a triangulated mesh. The smoothness parameter is fixed at \(\alpha=2\), and the replicate fields are conditionally independent given the shared hyperparameters. Thus, the Point model can be viewed as a Poisson regression model defined directly at the observed locations. 
Its main advantage is simplicity, since no numerical integration is required in the likelihood.

\subsection{Spatial model}
The second alternative interprets the observed count as an aggregated total over a planar spatial region. 
Let \(\{A_i^{S}\}_{i=1}^{M}\) denote the Voronoi cells induced by the unique crime locations in the planar study region \(D\), and let \(Y_{ir}\) be the count in cell \(A_i^{S}\) for replicate \(r\). 
The Spatial model is defined as
\begin{equation}
Y_{ir}\mid \mu_{ir}^{\mathrm{Spatial}}
\sim
\mathrm{Poisson}\!\left(\mu_{ir}^{\mathrm{Spatial}}\right),
\qquad
\mu_{ir}^{\mathrm{Spatial}}
=
\int_{A_i^{S}}
\exp\{\eta_r^{\mathrm{Spatial}}(s)\}\,ds.
\label{eq:mu_spatial_replicate}
\end{equation}
The latent predictor has the same form as the previous models,
\begin{equation*}
\eta_r^{\mathrm{Spatial}}(s)
=
\beta_0
+
\sum_{k=1}^{p}\beta_k z_k^{\mathrm{Spatial}}(s)
+
u_r^{\mathrm{Spatial}}(s),
\end{equation*}
where $z_k^{\mathrm{Spatial}}(s)
=
\exp\left\{-d_k^{\mathrm{Spatial}}(s)\right\}$ is defined through the Euclidean distance \(d_k^{\mathrm{Spatial}}(s)\) from \(s\) to the nearest amenity of type \(k\), and \(u_r^{\mathrm{Spatial}}(\cdot)\) is a zero-mean Gaussian Whittle--Mat\'ern field on \(D\) as introduced in \eqref{eq:spde_euclidean} with $\alpha = 2$, represented through the SPDE approach on a triangulated mesh.

In practice, the integral in \eqref{eq:mu_spatial_replicate} is approximated numerically. We construct planar integration points within each cell as described in the appendix. Let \(s_{ij}^{S}\) denote the \(j\)th integration point in cell \(A_i^{S}\), let \(w_{ij}^{S}\) be the quadrature weight associated with \(s_{ij}^{S}\), and let \(m_i\) be the number of integration points in \(A_i^{S}\), then
\begin{equation*}
\mu_{ir}^{\mathrm{Spatial}}
\approx
\sum_{j=1}^{m_i}
w_{ij}^{S}
\exp\{\eta_r^{\mathrm{Spatial}}(s_{ij}^{S})\}.
\label{eq:mu_spatial_quad_new}
\end{equation*}

\section{Simulation study}
\label{sec6}
We conducted a simulation study to compare the proposed metric-graph aggregated model (MG) with a planar aggregated model (Spatial). The main goal was to assess whether explicitly preserving the road-network geometry improves inference for aggregated count data generated on a street network. In all scenarios, the data were generated on the metric graph, so the MG model is correctly specified with respect to the latent spatial support, whereas the Spatial model serves as a misspecified benchmark that ignores the network constraint at the latent-process level.

\subsection{Simulation design}
\label{subsec:sim_design}
\begin{figure}[t]
    \centering
    \includegraphics[width=\textwidth]{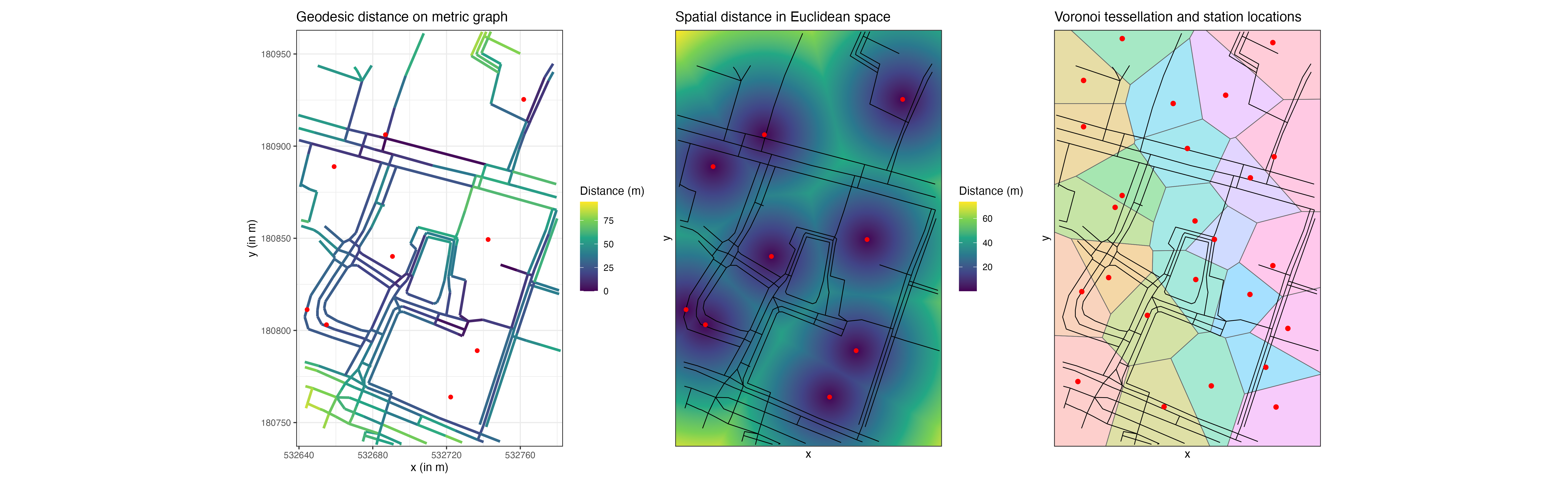}
    \caption{Illustration of the distance covariates and block construction used in the simulation study. The covariate defined by geodesic distance along the road network on the metric graph, corresponding to the MG model (left), the covariate defined by Euclidean distance in planar space, corresponding to the Spatial model (middle), and the Voronoi tessellation induced by the station locations used to construct the block partition (right). Red points indicate bar locations used as sources (left and middle panels) and the station locations (right panel).}
    \label{fig:sim_covariates_blocks}
\end{figure}

We considered a small street-network subregion in London, built from OpenStreetMap road data,  and represented it as a metric graph $\Gamma$ (see Figure~\ref{fig:sim_covariates_blocks}). Integration regions were constructed by first generating approximately uniformly distributed stations over the study window, and then computing a Voronoi tessellation. We intersect the Voronoi boundaries with graph edges to obtain intersection points and insert these points into the graph, thereby splitting original edges into smaller line segments. %Each resulting segment is assigned a \emph{block} label by mapping its midpoint to the nearest station. 
For each split segment, we generate numerical integration points and weights using Simpson's rule. %We collect these locations and weights in an \texttt{ips} data structure. 

For each replicate \(r\), we first generated a replicate latent Gaussian field \(u_r(s)\) on \(\Gamma\), using the Whittle--Mat\'ern model \eqref{eq:spde_metric_graph}, with $\alpha=1$,  \(\sigma=1\) and \(\rho = 1\,\mathrm{km}\). The latent log-intensity was specified as in \eqref{eq:mg_eta_replicate} with the covariate term restricted to a single distance-based covariate $z(s)=\exp\{-d(s)\}$ where \(d(s)\) is the geodesic distance on the road network from \(s\in\Gamma\) to the nearest bar location. For each block \(A_i\), the Poisson mean was defined by the integrated intensity in \eqref{eq:mg_mean_integral} and approximated using the quadrature scheme in \eqref{eq:mg_mean_quad}. 
Thus, the simulated dataset was generated from the MG model, with the covariate distance, latent Gaussian field, and aggregation structure all defined on the road network. The same simulated counts were subsequently fitted using two aggregated models: the proposed MG aggregated model and the Spatial aggregated model. In the Spatial model fit, the response was represented over planar observation blocks in \(D\subset\mathbb{R}^2\), the covariate distances were computed using Euclidean distance, and the Gaussian field was specified using the  SPDE formulation \eqref{eq:spde_euclidean} with \(\alpha=2\).

The baseline parameter setting was $n_{\mathrm{rep}} = 20$, $h = 0.00015$, $\beta_0 = 1$, $\beta = 10$, $n_{\mathrm{record}} = 5$, and $n_{\mathrm{sub}} = 5$. Here, $n_{\mathrm{rep}}$ denotes the number of independent replicates, $h$ denotes the integration step size on the metric graph and therefore controls the density of integration points for the MG model, $\beta_0$ is the intercept, $\beta$ is the covariate coefficient, $n_{\mathrm{record}}$ denotes the number of aggregation regions, and $n_{\mathrm{sub}}$ controls the numerical integration resolution for the Spatial model. The experimental settings are summarized in Table~\ref{tab:parameters}. We varied one factor at a time around the baseline setting. The values of $h$ and $n_{\mathrm{sub}}$ were chosen so that the numbers of integration points in the MG and Spatial models were of comparable magnitude. The corresponding values are summarized in Table~\ref{tab:integration_points}.

\begin{table}[t]
\centering
\caption{Baseline parameter settings and variation ranges.}
\label{tab:parameters}
\begin{tabular}{c c c}
\hline
Parameter & Baseline & Values considered \\
\hline
$n_{\mathrm{rep}}$ & 20 & $\{1,5,10,15,20\}$ \\
$\beta_0$ & 1 & $\{1,2,3,4,5\}$ \\
% $\beta$ & 10 & $\{1,5,10,15,20\}$ \\
$n_{\mathrm{record}}$ & 5 & $\{5,10,15,20\}$ \\
$n_{\mathrm{sub}}$ & 5 & $\{3,4,5,6,7\}$ \\
$h$ & $1.5\times 10^{-4}$ & $\{3.5,2.2,1.5,1.11,0.85\}\times 10^{-4}$ \\
%\hline
%$\sigma$ & 1 & -- \\
%$\rho$ & 1 & -- \\
%$\nu$ & 0.5 & -- \\
\hline
\end{tabular}
\end{table}

\begin{table}[ht]
\centering
\caption{Number of integration points for the MG and Spatial models}
\label{tab:integration_points}
\begin{tabular}{c c c c}
\hline
$n_{\mathrm{sub}}$ & $h$ & Planar points & MG points \\
\hline
3 & 0.00035  & 9559  & 9871  \\
4 & 0.00022  & 14934 & 15321 \\
5 & 0.00015  & 21507 & 22183 \\
6 & 0.000111 & 29269 & 29769 \\
7 & 0.000085 & 38233 & 38687 \\
\hline
\end{tabular}
\end{table}

\subsection{Simulation results}
\label{subsec:sim_results}
\begin{figure}[t]
  \centering

  \begin{subfigure}[t]{0.49\textwidth}
    \centering
    \includegraphics[width=\textwidth]{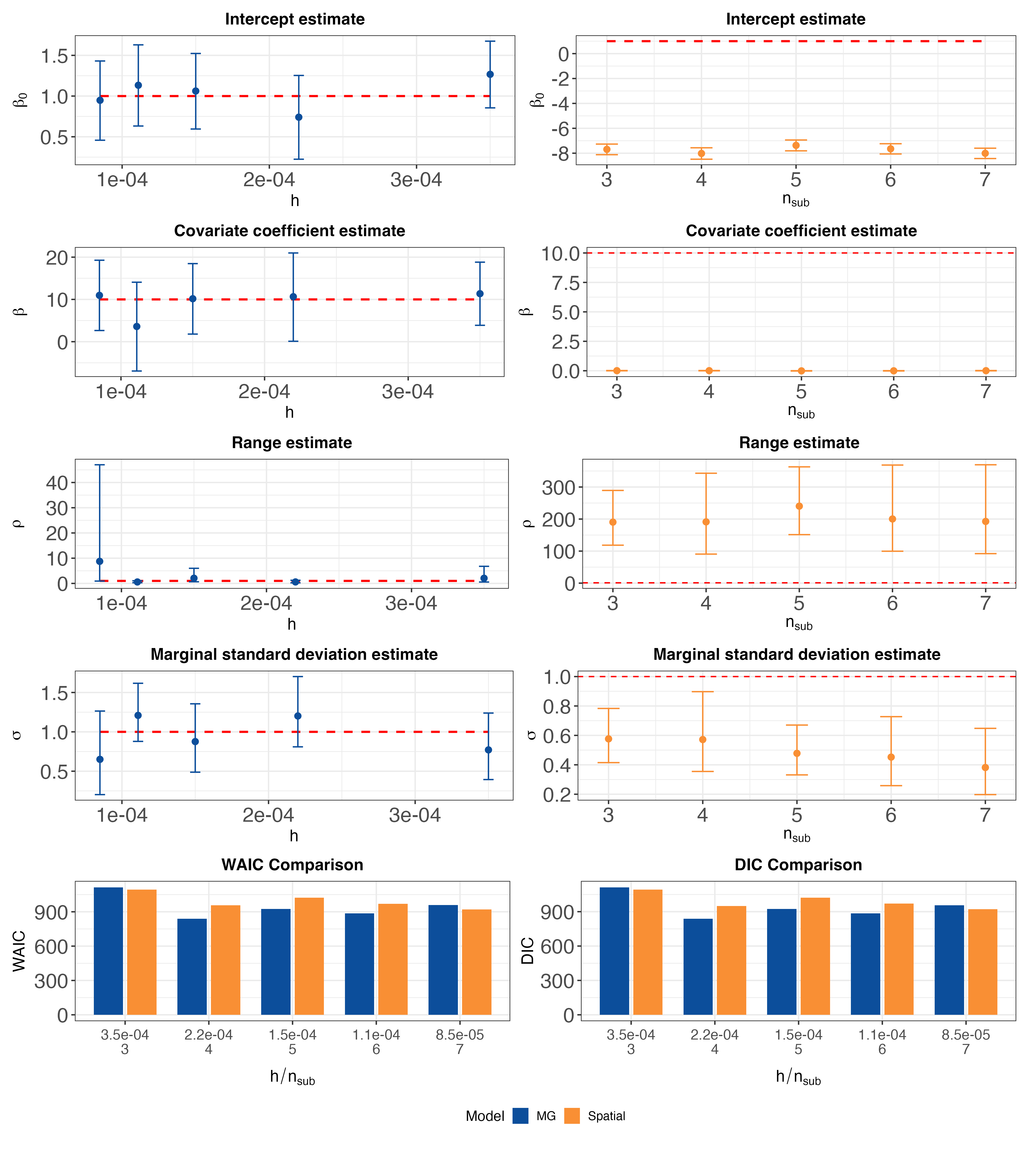}
    \caption{Varying $h$ / $n_{sub}$}
    \label{fig:poisson_sim_step}
  \end{subfigure}
  \hfill
  \begin{subfigure}[t]{0.49\textwidth}
    \centering
    \includegraphics[width=\textwidth]{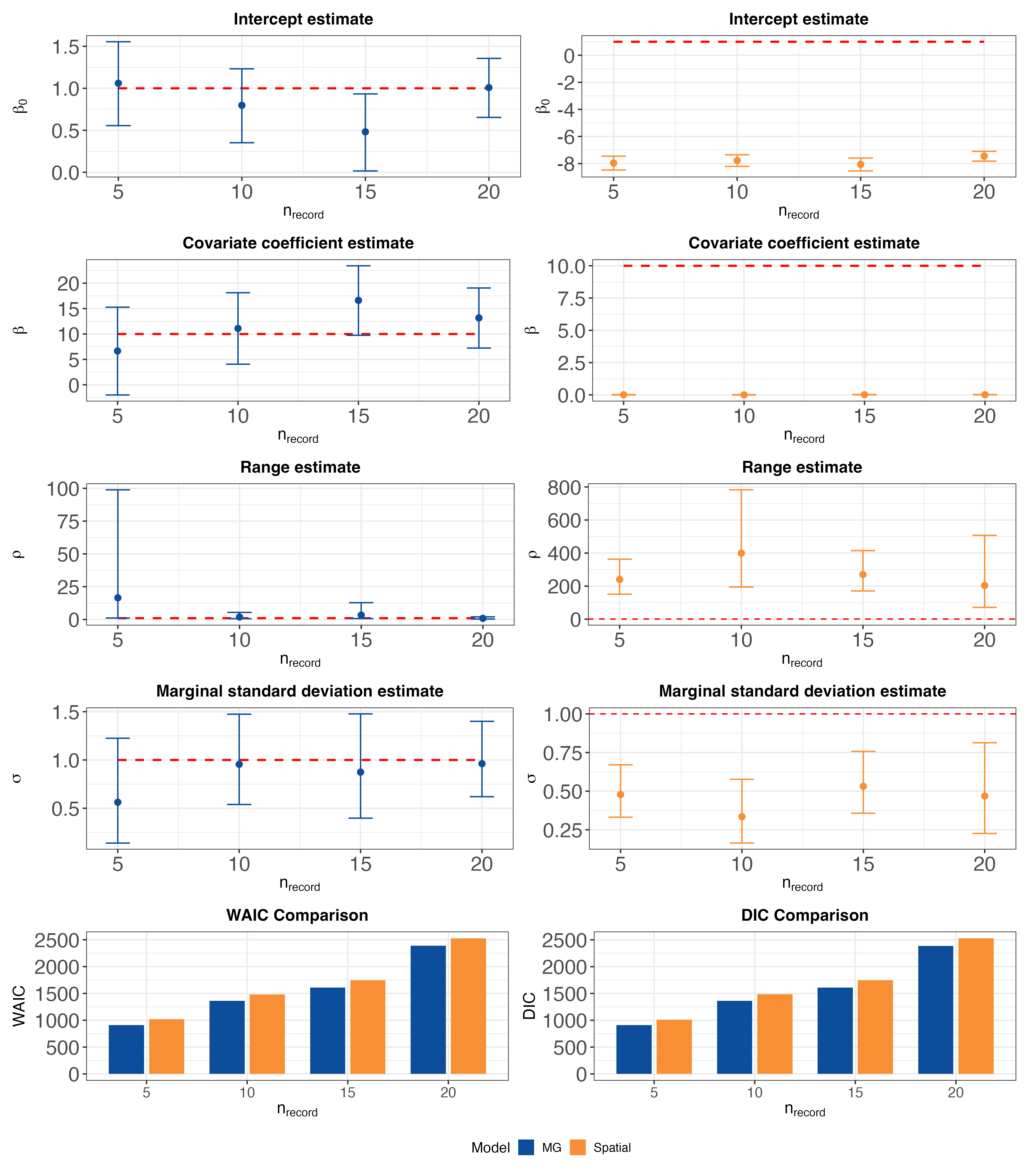}
    \caption{Varying $n_{record}$}
    \label{fig:poisson_sim_nrecord}
  \end{subfigure}

  \vspace{0.6em}

  \begin{subfigure}[t]{0.49\textwidth}
    \centering
    \includegraphics[width=\textwidth]{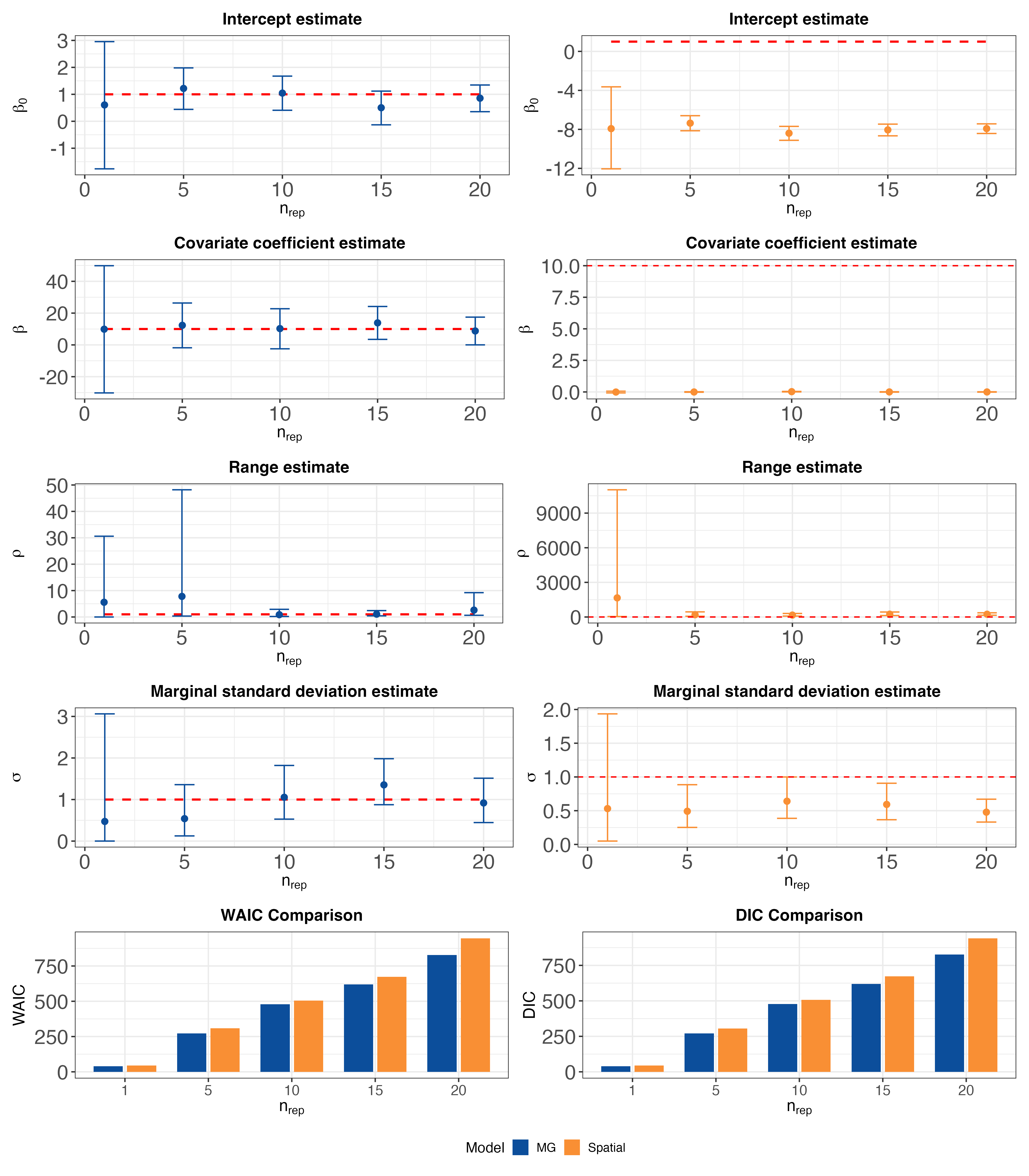}
    \caption{Varying \(n_{\mathrm{rep}}\).}
    \label{fig:poisson_sim_nrep}
  \end{subfigure}
  \hfill
  \begin{subfigure}[t]{0.49\textwidth}
    \centering
    \includegraphics[width=\textwidth]{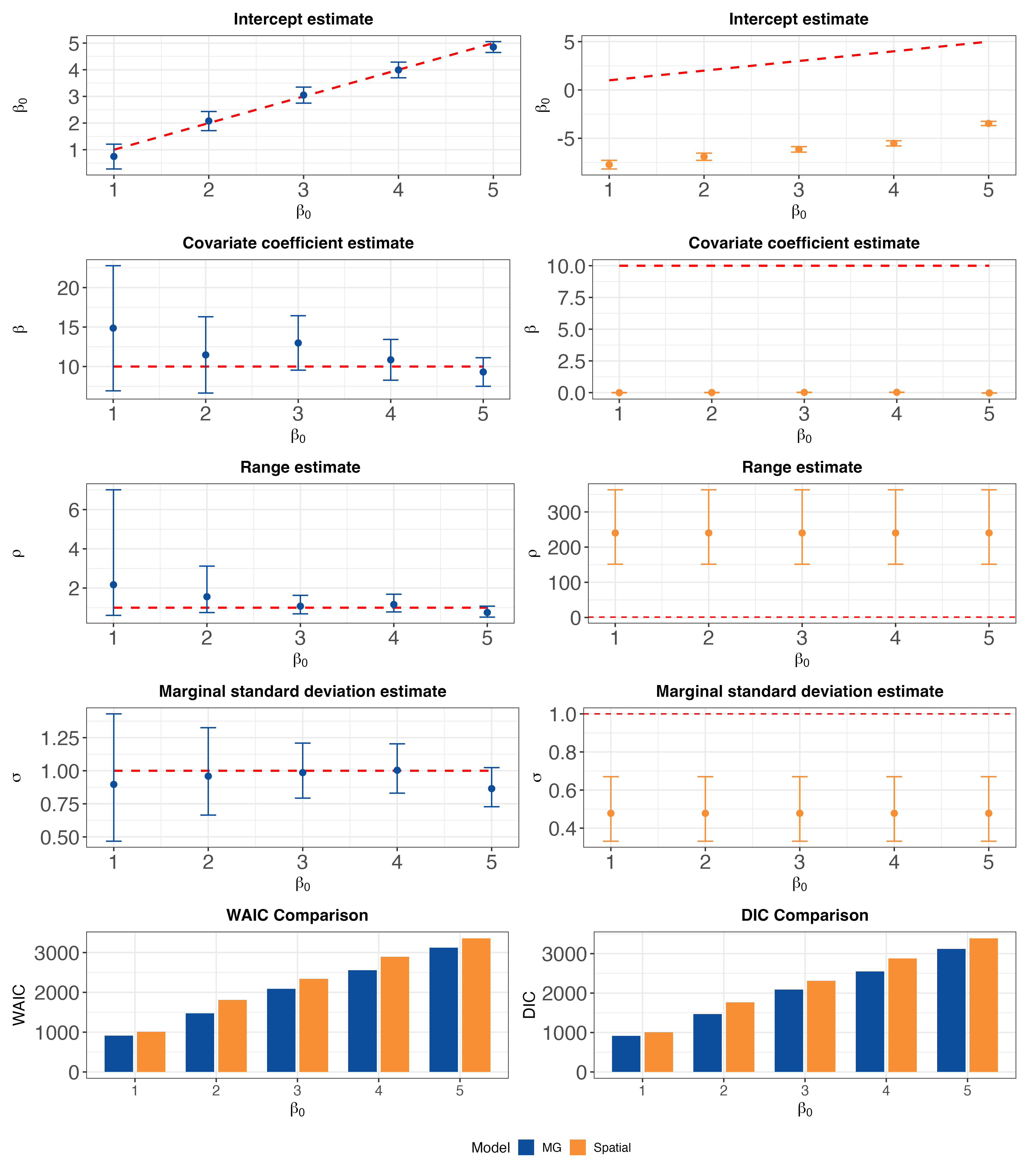}
    \caption{Varying \(\beta_0\).}
    \label{fig:poisson_sim_beta0}
  \end{subfigure}
  \caption{Simulation results under different integration resolutions, sample sizes, temporal replicates, and baseline intensities for the MG model (blue) and Spatial model (yellow). Red  lines indicate the true parameter values. Smaller WAIC and DIC values indicate better fit.}
  \label{fig:poisson_sim}
\end{figure}

The results are shown in Figure \ref{fig:poisson_sim}.
Under different $h$ values, the MG model remains broadly stable in its estimation of the fixed effects and spatial parameters, suggesting that the proposed method is not highly sensitive to the choice of integration resolution. 
A finer integration grid mainly leads to slightly narrower credible intervals and somewhat reduced variability in range parameter without changing the overall conclusions. In terms of computational scale, decreasing $h$ from 0.00035 to 0.000111 increases the number of integration points from 9871 to 38687. For the Spatial model, increasing $n_{\mathrm{sub}}$ from 3 to 7 increases the number of integration points from 9559 to 38233, but the overall bias pattern in the parameter estimates remains essentially unchanged. 
% n_record
The results also show that increasing the number of records improves the stability of the MG model. For the intercept, the MG model generally recovers the true level well and becomes more stable as the amount of information increases, whereas the Spatial model systematically underestimates the overall intensity. 
% beta
For the covariate coefficient \(\beta\), the MG model is able to reflect the true effect and track its changes across scenarios, while the Spatial model shrinks the estimate strongly toward zero in nearly all cases. This difference is especially important because the covariate is defined through shortest-path distance on the road network; once this network-based distance is replaced by Euclidean distance in the planar model, much of the regression signal is lost. For the range parameter, the MG model still shows some variability in more difficult settings, but its estimates remain substantially closer to the truth, whereas the Spatial model almost always overestimates the spatial range. 
For the marginal standard deviation \(\sigma\), the MG model typically stays near the true value, while the Spatial model tends to underestimate local spatial variation, suggesting that it redistributes variability incorrectly across the fixed and spatial components.
% n_rep
When the number of temporal replicates increases, the MG model becomes more stable and its uncertainty intervals generally become narrower, reflecting the additional information provided by repeated observations. In contrast, the Spatial model remains biased even with more replicates, suggesting that the bias is caused by geometric misspecification rather than limited sample size. 
% beta0 & beta
When the true intercept \(\beta_0\) changes, the MG model tracks the increasing baseline intensity well, whereas the Spatial model remains systematically shifted downward. Similarly, when the true covariate coefficient \(\beta\) increases, the MG model captures the stronger covariate signal, while the Spatial model estimates the covariate effect close to zero across all settings (these results are omitted from the figure to save space).
Overall, the advantage of the MG model is robust across changes in integration resolution, sample size, replicate number, baseline intensity, and covariate strength. 
This advantage is not simply due to numerical integration resolution or sample size, but comes from matching the spatial support of the model to the network-constrained data-generating mechanism. For the Spatial model, even increasing $n_{sub}$ and substantially enlarging the number of integration points does not remove the systematic bias caused by misspecification of the geometric support. Finally, the WAIC and DIC comparisons further confirm that the MG model provides the better overall fit in almost all scenarios.

\section{Application on the City of London}\label{sec7}
We now study the publicly available street-level crime data from \texttt{data.police.uk} introduced in Section \ref{sec2}. The aim is to compare crime risk modelling results under different assumptions on the underlying spatial support. Since the list of snap points in the UK Police data was updated in 2022, we use crime data from 2023 to 2025 to ensure that the location data are consistent with the updated snap-point system. UK Police also notes that, due to inconsistent geocoding policies across police forces, the provided location data may not be fully accurate or consistent. To account for potential coordinate errors and location uncertainty, points within 30 metres of each other are merged into a single location, thereby reducing duplicate or near-duplicate locations caused by geocoding errors. Let $\mathbf{s}_1,\dots,\mathbf{s}_M$ denote the set of unique station locations observed during the three-year period after merging. Based on these locations, we construct a planar Voronoi tessellation of the study region, where each Voronoi cell consists of all points in the plane that are closer to a given location $\mathbf{s}_i$ than to any other location. For the Spatial and MG models, these Voronoi cells are used to define the integration regions. In the Point model, each observation is associated directly with its corresponding location $\mathbf{s}_i$, and therefore no further spatial aggregation is required.

\subsection{Construct Integration Region}
\begin{figure}[t]
  \centering
  \includegraphics[width=1\linewidth]{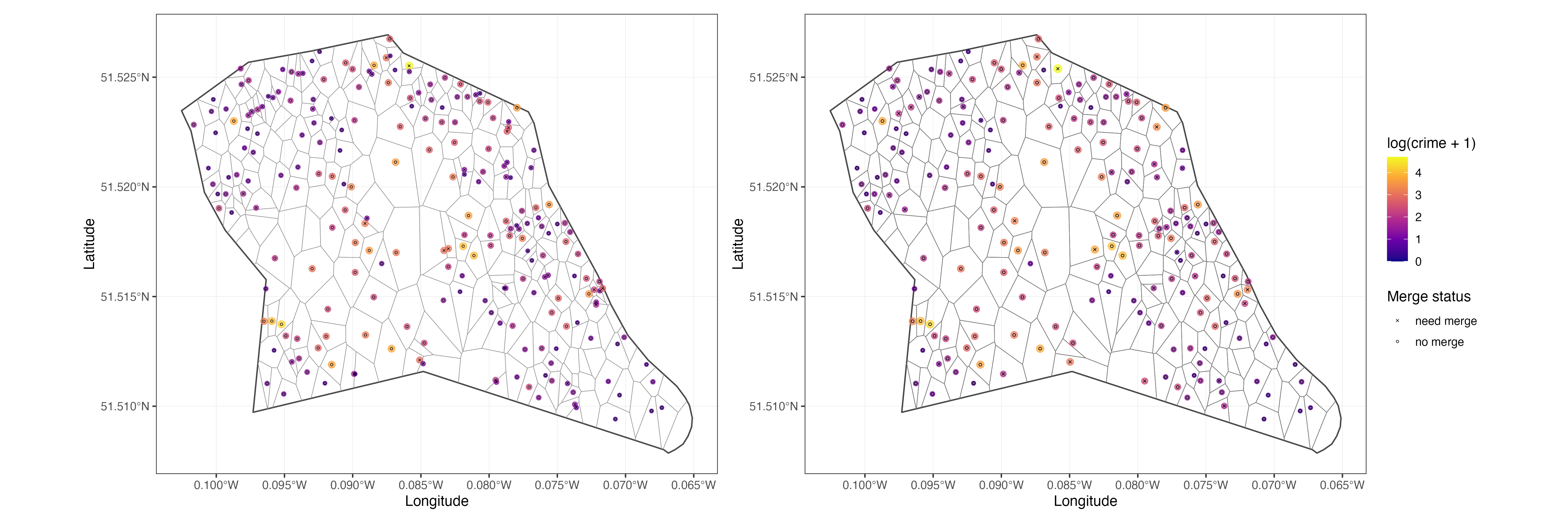}
  \caption{Planar Voronoi integration regions constructed from crime \textit{snap points} (left) and from station locations after applying the 30-metre merging threshold (right). 
  In both panels, Voronoi cells are constructed over the respective point configurations. The colour and size of points represent $\log(1+\mathrm{crime})$, and symbols indicate whether a location is affected by the 30-metre merging rule. A common legend is used for both panels.}
  \label{fig:voronoi_integration_regions}
\end{figure}

To illustrate the effect of the 30-metre distance threshold on the construction of integration regions, we construct planar Voronoi tessellations based on both the original crime \textit{snap points} and the merged station locations. Figure~\ref{fig:voronoi_integration_regions} shows the Voronoi integration regions generated from the original crime \textit{snap points}, where each cell corresponds to one original location. Since the geocoded locations provided by UK Police may contain positional errors and inconsistencies, nearby \textit{snap points} may represent the same or approximately the same underlying crime location. To reduce the influence of such location uncertainty on the definition of integration regions, points within 30 metres of each other are merged into a single station location, and the Voronoi tessellation is then reconstructed using the resulting unique locations, as also shown in Figure~\ref{fig:voronoi_integration_regions}. Compared with the original tessellation, the merged Voronoi tessellation reduces excessively small cells caused by near-duplicate locations and provides more stable integration regions.

\begin{figure}[t]
  \centering
  \includegraphics[width=1\linewidth]{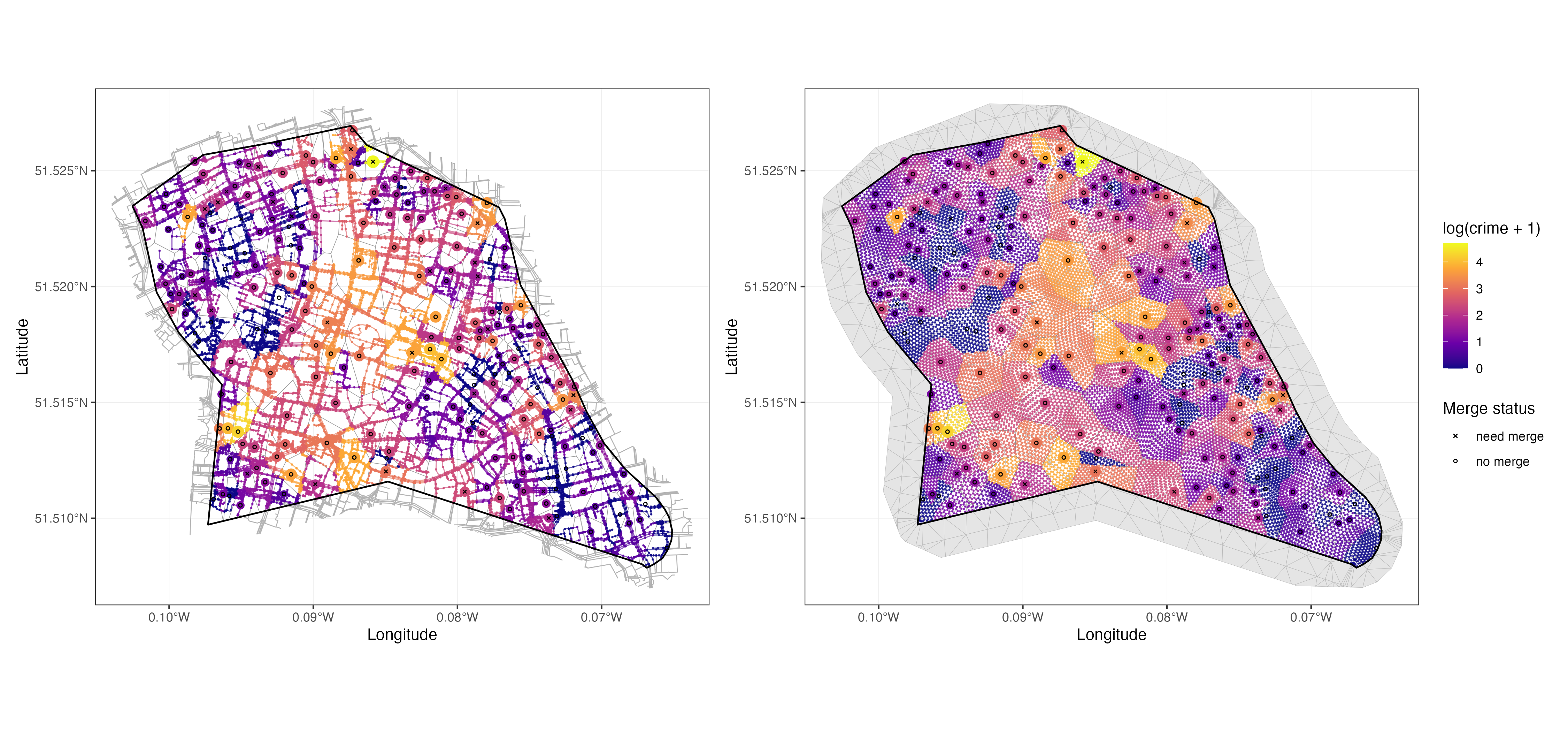}
  
  \caption{Comparison of numerical integration constructions under the Spatial model (left) and the metric graph (MG) model (right), based on merged Voronoi integration regions. In the Spatial model (left), integration points are placed on a mesh within the merged planar Voronoi regions. In the MG model (right), integration is performed on a metric graph, where Voronoi cells follow the road network and integration points are distributed along edges. In both panels, merged crime locations are shown, with point size and colour representing $\log(1+\mathrm{crime})$ in 2023. Voronoi boundaries are shown in grey and the study region boundary in black. A shared legend is used for consistent interpretation across panels.}
  
  \label{fig:spatial_vs_mg_merged_ips}
\end{figure}

For the Spatial model, the merged Voronoi cells are used as areal integration regions. To reduce potential boundary effects, the triangular mesh is constructed with an offset that provides a small buffer beyond the study boundary. Although the mesh extends outside the study region, the Voronoi integration regions remain restricted to the City of London. Numerical integration points are then generated within each Voronoi polygon using the mesh-based integration scheme. In this setting, the intensity function is integrated over two-dimensional areal units, and the expected count for each unit is obtained by summing the weighted contributions of the integration points inside the corresponding Voronoi cell. Figure~\ref{fig:spatial_vs_mg_merged_ips} illustrates the spatial integration structure based on the merged station locations. 

For the MG model, the integration regions are further constrained by the road network. 
To avoid artificially truncating streets that remain connected immediately outside the study region, the metric graph is extended slightly beyond the City of London boundary. The integration regions remain restricted to the original study boundary. We intersect the Voronoi boundaries generated from the merged crime locations with the metric graph. All intersection points are added to the graph as new vertices, thereby splitting the original road edges into a refined set of shorter line segments $\{e_k\}_{k=1}^K$. For each refined segment $e_k$, numerical integration points are constructed along the road network using the Simpson-based integration scheme described in Section~\ref{sec3}. The midpoint of each refined segment is used as its representative location, and the segment is assigned to the nearest merged crime location. This produces a Voronoi-induced partition of the metric graph, where each block represents the linear domain of influence of one merged crime location on the road network. For the replicated data structure, the same set of MG integration points is duplicated $R$ times, where $R$ denotes the number of years. Within replicate $r$, the crime counts for year $r$ are mapped to the corresponding integration blocks according to their year and location. If a location appears in other years but has no observed crime event in year $r$, its count is set to zero for that year. Figure~\ref{fig:spatial_vs_mg_merged_ips} also shows the Voronoi-induced aggregation structure constructed on the metric graph. 

\subsection{Covariates}
Crime occurrence is often influenced by the surrounding street environment and the distribution of urban amenities. Previous studies have shown systematic associations between different types of facilities and street crime. For example, \cite{wullenweber2024crime} reported that robbery is more likely to occur near entertainment venues with poor nighttime lighting and high pedestrian density, such as bars and nightclubs; crime hotspots are also often found around ATMs and bus stations. \cite{sidebottom2010bag} found that crime tends to cluster in and around bars. \cite{bowers2014risky} also found that retail outlets and commercial facilities are associated with higher levels of street theft. Motivated by this literature, we construct a set of distance-based covariates derived from urban amenities, including bars, nightclubs, bus stations, subway entrances, supermarkets, and convenience stores. For each facility type, we first extract its locations within the study region and then compute, on the road network, the shortest-path distance from each model support point to the nearest facility of that type. These distances are subsequently transformed using the exponential decay function introduced in Section \ref{sec4}.

\subsection{Results}
We use the models introduced in Sections~\ref{sec4} and~\ref{sec5}. The metric-graph aggregated model (MG) is used as the main model, while the Point and Spatial models are fitted as planar alternatives for comparison. All three models use the same set of fixed covariates in the latent log-intensity, namely distance-based covariates for bars, nightclubs, bus stations, subway entrances, supermarkets, and convenience stores.

\begin{figure}[t]
  \centering
\includegraphics[width=0.8\linewidth]{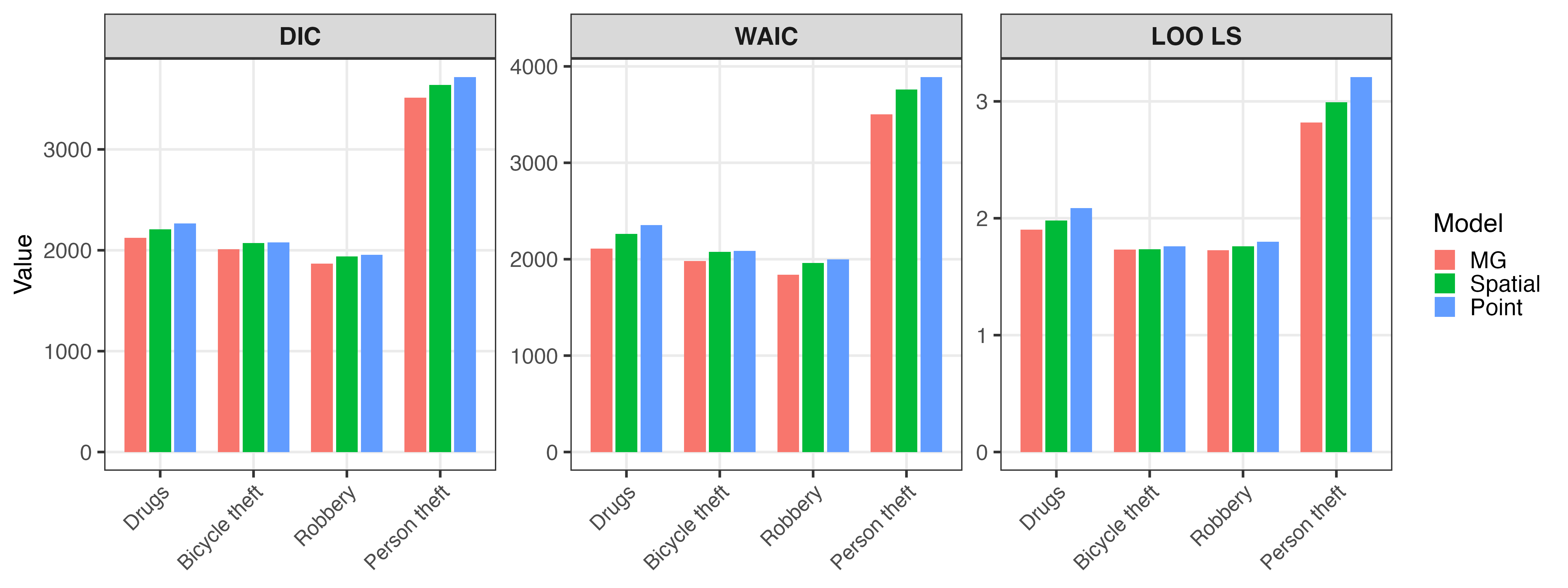}
  \caption{Model comparison across crime types using DIC, WAIC and leave-one-out crossvalidation log-score.}
  \label{fig:model_compare}
\end{figure}

Model comparison results are shown in Figure~\ref{fig:model_compare}. Overall, the MG model provides the best model fit and predictive performance across the four crime types. Model performance is evaluated using DIC, WAIC, and the leave-one-out log score (LOO LS), where $\mathrm{LOO\ LS}=-\frac{1}{n}\sum_{i=1}^{n}\log p(y_i \mid y_{-i})$, with $p(y_i \mid y_{-i})$ denoting the leave-one-out posterior predictive probability of observation $y_i$ obtained using all observations except $y_i$. Therefore, a smaller LOO LS indicates better out-of-sample predictive performance. For Drugs, Bicycle theft, Robbery, and Theft from the person, the MG model generally has lower DIC, WAIC, and LOO LS values than the Spatial and Point models. These results indicate that the metric graph-based modelling framework can more effectively describe the spatial distribution of street-level crime while also providing better leave-one-out predictive performance. The result also indicates that crime risk does not simply diffuse continuously over two-dimensional geographic space, but is more likely to be shaped by the road network, street-segment structure, and intersection connectivity. By mapping crime events onto the road network and introducing an SPDE spatial random effect on the metric graph, the MG model is better able to capture the clustering and propagation of street crime along the network structure.

\begin{figure}[t]
  \centering
  \includegraphics[width=1\linewidth]{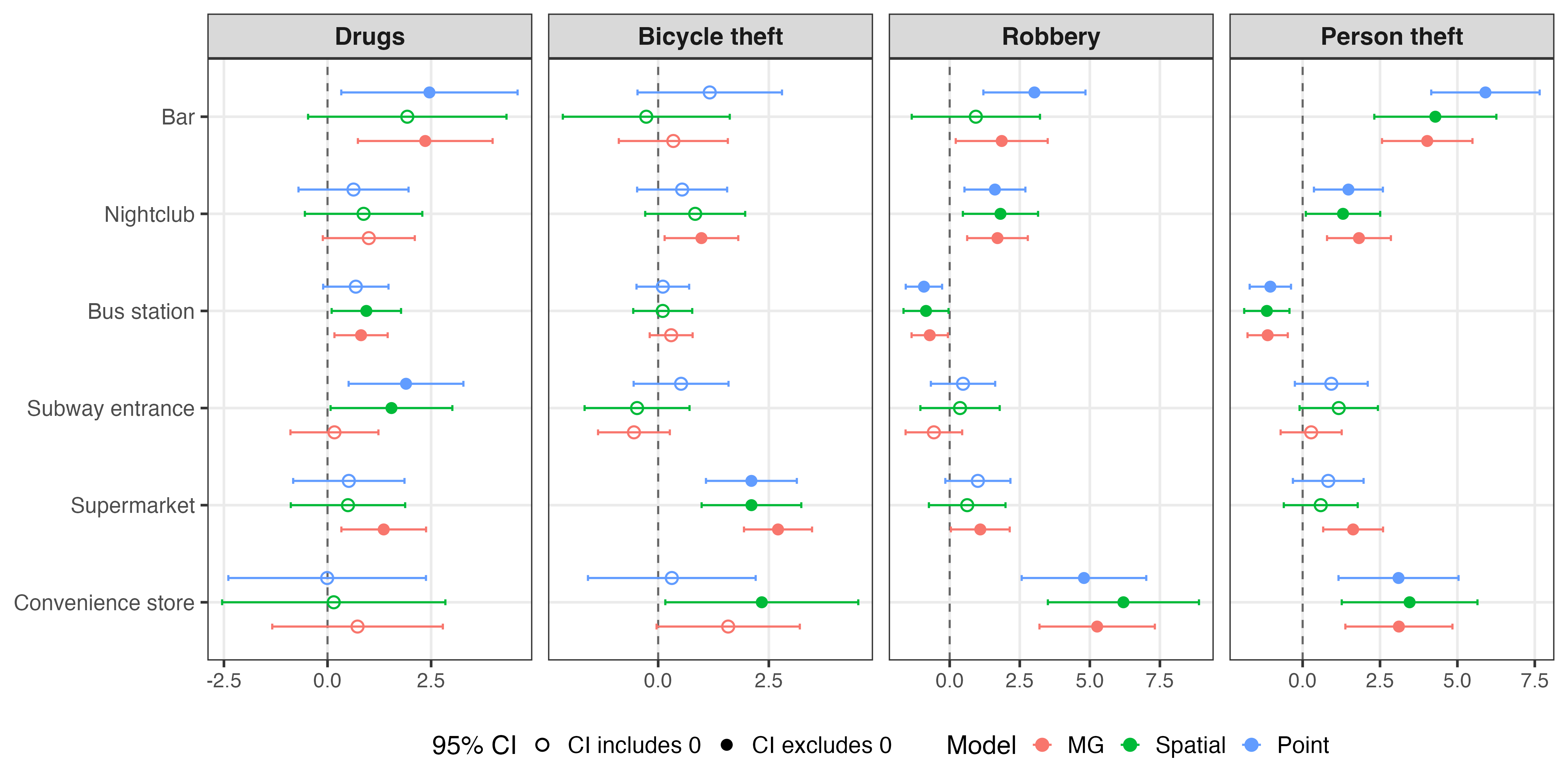}
  \caption{Posterior means and 95\% credible intervals of fixed effects for selected crime types under the Point, Spatial, and MG models. Points show posterior means, horizontal intervals show 95\% credible intervals, and the vertical dashed line indicates zero. Solid horizontal intervals indicate that the 95\% credible interval does not include zero.}
  \label{fig:FE}
\end{figure}

Figure~\ref{fig:FE} presents the posterior means and 95\% credible intervals of the fixed effects under the Point, Spatial, and MG models for the four crime types. The points represent posterior means, the horizontal intervals represent 95\% credible intervals, and the vertical dashed line indicates zero. If a credible interval includes zero, it is shown as a dashed interval. The fixed-effect results show that the effects of environmental covariates differ across crime types. 
Under the MG model, supermarkets show positive effects for Drugs, Bicycle theft, Robbery, and Theft from the person, suggesting that areas around supermarkets may be associated with higher crime risk. This may be because supermarkets attract regular pedestrian traffic.
Nightclubs show positive effects for Bicycle theft, Robbery, and Theft from the person, while their effect on Drugs is not clearly different from zero. This suggests that night-time activity areas may be associated with higher risks for certain crime types, possibly due to crowding, late-night mobility, and increased opportunities for crime.
For subway entrances, the effects are uncertain across all four crime types under the MG model.
Convenience stores show positive effects for Robbery and Theft from the person, suggesting that everyday retail locations may increase pedestrian activity and target exposure. For Drugs and Bicycle theft, the effects are uncertain.
Bars show positive effects for Drugs, Robbery, and Theft from the person, while their effect on Bicycle theft is uncertain. This suggests that bars and the activities around them may be associated with increased risks for specific crime types.
The effects of bus stations also vary across crime types. Under the MG model, bus stations show a positive effect for Drugs and negative effects for Robbery and Theft from the person.
Overall, under the MG model, supermarkets, nightclubs, convenience stores, and bars are associated with elevated risks for specific crime types.

The fixed-effect estimates are generally similar across the three models, with a few notable differences. For example, for Drugs, subway entrances have significant positive effects under the Point and Spatial models but are insignificant under the MG model, whereas supermarkets have a significant positive effect under the MG model but are insignificant under the other two models. For Bicycle theft, nightclubs show a significant positive effect only under the MG model. In addition, for Robbery and Theft from the person, supermarkets have significant positive effects under the MG model, while their credible intervals include zero under the Point and Spatial models. These differences show that the choice of spatial model and the assumed spatial support can affect both the fixed-effect estimates and their associated uncertainty, and may therefore influence conclusions about the relationships between environmental covariates and crime risk.

\section{Conclusion}
\label{sec8}
We have developed an LGCP framework for modelling street-level crime data on a metric graph when the observed event locations are spatially obfuscated. The main contribution is to align the statistical model with both the geometry of the crime process and the support induced by the data-release and anonymisation mechanism. Instead of treating the released coordinates as exact crime locations, we represent each observation through an aggregated support on the street network and model the resulting counts using an area-aggregated LGCP. The simulation study shows that preserving the metric-graph support is important when the underlying process is constrained to a road network. When data are generated on the network, the proposed MG model gives more reliable parameter estimates than the planar spatial alternative. In particular, the MG model better recovers the intercept, the covariate effect, and the spatial dependence parameters, while the planar model tends to underestimate the covariate effect and distort the spatial range. These results indicate that increasing the numerical integration resolution in a misspecified planar model cannot fully correct the bias caused by using the wrong spatial support. The application to City of London crime data further supports the usefulness of the proposed framework. Across the selected crime types, the MG model provides better overall fit than both the point-based and planar aggregated models. This suggests that street crime risk is not adequately represented as a process evolving over ordinary two-dimensional space, but is more naturally described as a process constrained by the road network. By combining the network geometry with an aggregated likelihood, the MG model is able to account for both the linear structure of streets and the uncertainty introduced by location anonymisation. 

The fixed-effect results also show that the relationships between environmental amenities and crime risk vary across crime types. Under the MG model, supermarkets show positive effects across all four crime types. Nightclubs are positively associated with Bicycle theft, Robbery, and Theft from the person, while bars show positive effects for Drugs, Robbery, and Theft from the person. Convenience stores are mainly associated with Robbery and Theft from the person. Although the three models generally produce fixed-effect estimates with similar directions, several differences in statistical significance are observed. This indicates that the choice of spatial support can influence not only overall model fit but also inference on the relationships between environmental covariates and crime risk. Overall, these findings suggest that different crime types are associated with different urban activity patterns and that explicitly accounting for the street-network support can provide a more appropriate representation of local crime risk.

\begin{appendix}
\section{Construction of spatial integration points}
\label{supp:spatial_ips}

 In the two-dimensional spatial setting, let the study region be \(D \subset \mathbb{R}^2\). Denote the set of mesh triangles by
 $\{T_k\}_{k=1}^K$,
 where the three vertices of triangle \(T_k\) are
 $v_{k1},\; v_{k2},\; v_{k3}\in\mathbb{R}^2$.
 For an aggregated observation region \(A_i^{S}\), the expected count is
 \[
 \mu_i=\int_{A_i^{S}}\exp\{\eta(s)\}\,ds.
 \]
 The integration points are first constructed on a reference triangle and then mapped to each mesh triangle using the \texttt{fmesher} \citep{fmesher} package. For a given subdivision parameter \(n_{\mathrm{sub}}\), define
 $n_B=(n_{\mathrm{sub}}+1)^2$.
 Then construct the one-dimensional sequence $b = \{b_r\}_{r=0}^{n_{\mathrm{sub}}}$ with 
 \[
 b_r=\frac{\nicefrac1{3}+r}{n_{\mathrm{sub}}+1},\qquad r=0,1,\dots,n_{\mathrm{sub}},
 \]
 and form the two-dimensional grid
 $(b_1,b_2)\in b\times b$.
 For every point satisfying
 $b_1+b_2>1$,
 the following reflection is applied:
 $(b_1,b_2)\mapsto(1-b_2,\;1-b_1)$.
 After this transformation, all points satisfy
 $b_1\ge 0$, $b_2\ge 0$, and $b_1+b_2\le 1$.
 A third barycentric coordinate is then constructed as 
 $(\lambda_1,\lambda_2,\lambda_3)
 =
 (1-b_1-b_2,\;b_1,\;b_2)$,
 so that
 $\lambda_1+\lambda_2+\lambda_3=1$, and $\lambda_r\ge 0$, for $r=1,2,3$.
 If an integration point has barycentric coordinates
 $(\lambda_1,\lambda_2,\lambda_3)$,
 then its physical location in triangle \(T_k\) is
 $u_{kj}
 =
 \lambda_1 v_{k1}+\lambda_2 v_{k2}+\lambda_3 v_{k3}$, for 
 $j=1,\dots,n_B$.
 Thus, each triangle \(T_k\) contains \(n_B\) integration points. Let \(|T_k|\) denote the area of triangle \(T_k\). Each integration point in \(T_k\) is assigned the same weight
 \[
 w_{kj}=\frac{|T_k|}{n_B},\qquad j=1,\dots,n_B.
 \]
 After obtaining the full set of integration points
 \[
 \{(u_{kj},w_{kj}): k=1,\dots,K,\; j=1,\dots,n_B\},
 \]
 they are reassigned to the aggregated integration regions according to their spatial locations. %In this paper, these regions are the Voronoi cells \(\{A_i^{S}\}_{i=1}^M\). 
 For each integration point \(u_{kj}\), if
 $u_{kj}\in A_i^{S}$,
 then the point and its weight \(w_{kj}\) are assigned to region \(A_i^{S}\). Consequently, %for any integrable function \(f(s)\),
 %\[
 %\int_{A_i^{S}} f(s)\,ds
 %\approx
 %\sum_{(k,j):\,u_{kj}\in A_i^{S}} w_{kj}f(u_{kj}).
 %\]
 for the LGCP model,
 \[
 \mu_i
 =
 \int_{A_i^{S}}\exp\{\eta(s)\}\,ds
 \approx
 \sum_{(k,j):\,u_{kj}\in A_i^{S}} w_{kj}\exp\{\eta(u_{kj})\}.
 \]
\end{appendix}

\bibliographystyle{imsart-nameyear} 
\bibliography{Ref2}

\end{document}